\documentclass[11pt,a4paper]{article}

\usepackage[utf8]{inputenc}
\usepackage[T1]{fontenc}
\usepackage{lmodern}
\usepackage[margin=2.5cm]{geometry}
\usepackage{amsmath,amssymb}
\usepackage{graphicx}
\usepackage{booktabs}
\usepackage{hyperref}
\usepackage{url}
\usepackage{natbib}
\usepackage{xcolor}
\usepackage{caption}
\usepackage{subcaption}
\usepackage{multirow}
\usepackage{tabularx}
\usepackage{authblk}
\usepackage{abstract}
\usepackage{float}

\hypersetup{
    colorlinks=true,
    linkcolor=blue!70!black,
    citecolor=blue!70!black,
    urlcolor=blue!70!black
}

\title{\textbf{AI-Driven Modular Services for Accessible Multilingual Education in Immersive Extended Reality Settings: Integrating Speech Processing, Translation, and Sign Language Rendering}}

\author[1]{N.~D.~Tantaroudas\thanks{Corresponding author: nikolaos.tantaroudas@iccs.gr}}
\author[2]{A.~J.~McCracken}
\author[3]{I.~Karachalios}
\author[4]{E.~Papatheou}

\affil[1]{Institute of Communications and Computer Systems (ICCS), Iroon Polytechneiou 9, 15773 Zografou, Athens, Greece}
\affil[2]{DASKALOS-APPS, 183 Rue de l'Abb\'{e} Griffon, 01960 P\'{e}ronnas, France}
\affil[3]{National Technical University of Athens, Leof. Alimou, Katechaki, Zografou 15772, Athens, Greece}
\affil[4]{Exeter Small-Scale Robotics Laboratory, Engineering Department, University of Exeter, Exeter EX4 4QF, UK}

\date{}

\begin{document}
\maketitle

% ============================================================
% ABSTRACT
% ============================================================
\begin{abstract}
\noindent\textbf{Background:} Extended Reality (XR) technologies hold considerable promise for revolutionising language education, yet existing platforms overwhelmingly overlook the accessibility requirements of deaf and hard-of-hearing learners. Most current solutions function within monolingual settings and do not provide integrated sign language or multimodal communication capabilities. A pressing demand exists for inclusive systems that cater to both deaf and hearing learners via cross-modal artificial intelligence (AI) services embedded within immersive environments.

\noindent\textbf{Methods:} This work introduces a modular platform that brings together six AI services: automatic speech recognition via OpenAI Whisper, multilingual translation through Meta NLLB, speech synthesis using AWS Polly, emotion classification with RoBERTa, dialogue summarisation via flan-t5-base-samsum, and International Sign (IS) rendering through Google MediaPipe. A corpus of 750 IS gesture recordings was processed to derive hand landmark coordinates, which were subsequently mapped onto three-dimensional avatar animations inside a Unity-based virtual reality (VR) environment running on Meta Quest~3 headsets. Validation comprised technical benchmarking of each AI component, including comparative assessments of speech synthesis providers and multilingual translation models (NLLB-200 and EuroLLM 1.7B variants), together with load testing to gauge system scalability.

\noindent\textbf{Results:} Technical evaluations confirmed the platform's suitability for real-time XR deployment. Speech synthesis benchmarking established that AWS Polly delivers the lowest latency (50--100\,ms time to first byte) at a competitive price point. The EuroLLM 1.7B Instruct variant attained a BLEU score of 84.34, surpassing NLLB's score of 79.25. Stress testing with 1{,}000 simulated concurrent users yielded average response times below 800\,ms with no critical failures. Avatar animation latency for IS gesture rendering consistently remained under 300\,ms.

\noindent\textbf{Conclusions:} These findings establish the viability of orchestrating cross-modal AI services within XR settings for accessible, multilingual language instruction. The modular design permits independent scaling and adaptation to varied educational contexts, providing a foundation for equitable learning solutions aligned with European Union digital accessibility goals.

\end{abstract}

% ============================================================
% PLAIN LANGUAGE SUMMARY
% ============================================================
\section*{Plain Language Summary}
Acquiring a new language is demanding, and this challenge is magnified for deaf individuals who depend on sign language. This research tackles the problem by developing a virtual reality (VR) learning space in which a digital three-dimensional character (avatar) can speak, translate, and perform sign language gestures in real time. The platform employs multiple AI tools in concert: one converts spoken language into text, another translates text across numerous languages, a third transforms text back into speech, and a fourth renders text as International Sign Language gestures performed by the avatar. Learners wear a VR headset and engage with the avatar inside a virtual classroom, where they can choose their preferred language and receive immediate translations in both spoken and signed modalities. The platform's technical performance was confirmed through benchmarking of AI translation models, speech synthesis services, and scalability assessments, verifying its fitness for real-time educational use. This research represents a stride toward building virtual learning spaces where language barriers and hearing limitations no longer obstruct access to education.

% ============================================================
% 1. INTRODUCTION
% ============================================================
\section{Introduction}
\label{sec:intro}

Foreign language learning has traditionally depended on structured pedagogical strategies, encompassing grammar-translation approaches, audio-lingual exercises, and immersion techniques~\citep{shliakhtina2023grammar, wu2023perceptions}. The grammar-translation paradigm, grounded in classical education, places emphasis on textual analysis and memorisation of grammatical rules, whereas audio-lingual methods centre on pattern repetition and habitual formation~\citep{shliakhtina2023grammar}. Immersion-based strategies seek to replicate natural language acquisition by situating learners within target-language contexts~\citep{wu2023perceptions}. Over recent decades, technological instruments such as language laboratories, multimedia applications, and computer-assisted instruction have supplemented these conventional approaches, allowing learners to practise independently~\citep{wu2023perceptions, divekar2021foreign}.

The advent of Extended Reality (XR), spanning Augmented Reality (AR), Virtual Reality (VR), and Mixed Reality (MR), has ushered in transformative opportunities for language education by providing immersive, context-rich simulations that transcend the constraints of traditional classroom environments~\citep{divekar2021foreign, tegoan2021application}. AR superimposes digital content onto the physical world, enhancing vocabulary acquisition through contextualised visual and auditory stimuli, while VR fully immerses learners in synthetic environments where they can participate in conversational practice with virtual interlocutors~\citep{tegoan2021application, panagiotidis2021virtual}. Recent investigations have shown that VR environments can bolster learners' speaking confidence and diminish anxiety by offering risk-free, repeatable practice scenarios~\citep{godwinJones2023presence, zhi2023extended}. Commercial solutions such as ImmerseMe VR~\citep{immerseme} enable learners to navigate realistic situations, such as ordering food in a restaurant or requesting directions, thereby fostering contextual learning, while AR applications like MondlyAR~\citep{mondlyar} exploit spatial anchoring to reinforce vocabulary retention.

Artificial Intelligence (AI) complements XR by driving adaptive learning systems, natural language processing (NLP), and real-time feedback mechanisms~\citep{garcia2024binding}. AI-driven 3D avatars within XR environments can provide personalised language instruction, support conversational practice, and adjust to individual learner requirements, generating a more engaging and effective educational experience~\citep{panagiotidis2021virtual, hartholt2019ubiquitous}. The convergence of AI and XR has been recognised as a pivotal enabler of immersive, learner-centred approaches to language acquisition, narrowing the gap between structured classroom exercises and authentic communicative practice~\citep{godwinJones2023presence, zhang2023concepts}.

Nonetheless, notable gaps persist in the accessibility and inclusivity of current XR-based language learning platforms~\citep{taborda2025engagement, zhang2023concepts}. Most existing systems are tailored for monolingual environments or narrowly scoped use cases, constraining their usefulness for multilingual and multicultural learners~\citep{taborda2025engagement}. More critically, these platforms largely disregard the needs of deaf and hard-of-hearing individuals, who depend on sign language for effective participation in language education~\citep{taborri2023use, strobel2023artificial}. Although progress has been made in AI-driven sign language recognition, the development of comprehensive text-to-sign translation systems remains hampered by insufficient annotated datasets, limited accuracy in gesture recognition, and the lack of real-time translation capabilities within XR settings~\citep{chaudhary2022signnet, rodriguezcorrea2023benefits}. These challenges are compounded by the diversity of sign languages across regions, with the majority of research focusing on American Sign Language (ASL) while largely neglecting International Sign (IS), a visual communication system employed by deaf individuals from varied linguistic backgrounds in international settings~\citep{eud2018position, yin2023gloss}.

Moreover, while AI-driven avatars hold substantial promise for natural, culturally attuned language instruction, they presently struggle to deliver precise interactions across different languages and cultural nuances, owing to challenges in NLP algorithms and the complexity of implementing diverse linguistic databases~\citep{taborri2023use, sylaiou2023use}. Robust pedagogical frameworks that steer the effective incorporation of XR technologies into language curricula are likewise required to ensure that technological innovation translates into meaningful educational outcomes~\citep{zhang2023concepts, hirzle2023when}.

The imperative for inclusive educational technologies is further highlighted by European Union policy frameworks, including the European Accessibility Act and the European Strategy for the Rights of Persons with Disabilities 2021--2030, which advocate for accessible digital services and equitable participation in education across member states. The EU's Digital Education Action Plan (2021--2027) specifically underlines the role of emerging technologies, including AI and immersive environments, in fostering inclusive and high-quality education. Despite these policy initiatives, a significant disparity remains between the accessibility ambitions articulated in EU frameworks and the practical capabilities of current XR-based educational platforms, particularly concerning support for sign language users and multilingual learners.

This paper addresses these gaps by presenting a comprehensive platform that unifies modular AI-driven services for accessible language education in immersive XR environments. Extending and building upon our earlier publications~\citep{tantaroudas2026xrsalento, tantaroudas2026business, tantaroudas2026career, tantaroudas2026interact, tantaroudas2026aiservices}, this manuscript provides an expanded review of the literature, describes the complete system architecture, presents quantitative benchmarking analyses of AI translation models and speech synthesis services, and examines the implications for equitable educational technology. The contributions of this work are summarised as follows:
\begin{itemize}
    \item[(a)] The design and implementation of a modular, interoperable framework that combines speech-to-text, text-to-speech, text-to-text translation, sentiment analysis, and IS translation within a unified XR platform;
    \item[(b)] The creation of an IS gesture dataset comprising 750 videos processed using Google MediaPipe for real-time avatar-driven sign language delivery;
    \item[(c)] Quantitative benchmarking of multilingual translation models (NLLB-200 vs.\ EuroLLM) and text-to-speech services; and
    \item[(d)] A scalability evaluation demonstrating robust performance under simulated high-demand conditions.
\end{itemize}

The remainder of this paper is structured as follows. Section~\ref{sec:related} presents the related work spanning XR for education, AI-driven translation, sign language processing, and session summarisation. Section~\ref{sec:methodology} details the methodology and system architecture. Section~\ref{sec:results} presents the results, encompassing AI service implementations and benchmarking analyses. Section~\ref{sec:discussion} discusses the findings and their implications, and Section~\ref{sec:conclusions} concludes the paper with directions for future research.

% ============================================================
% 2. RELATED WORK
% ============================================================
\section{Related Work}
\label{sec:related}

\subsection{Extended Reality for Language Education}
\label{sec:related_xr}

The deployment of XR technologies in education has expanded substantially, propelled by the recognition that immersive environments can enhance engagement, motivation, and learning retention~\citep{divekar2021foreign, tegoan2021application}. In the domain of language learning, VR offers a distinctive opportunity to place learners within authentic communicative contexts, enabling them to practise speaking, listening, and interacting in a target language without the social pressures inherent in real-world interactions~\citep{godwinJones2023presence, zhi2023extended}. Divekar et al.~\citep{divekar2021foreign} demonstrated that foreign language acquisition systems merging AI with XR can substantially improve learner outcomes by delivering contextualised, adaptive interactions. Tegoan et al.~\citep{tegoan2021application} conducted a systematic review of XR applications for language instruction, concluding that immersive technologies provide distinct advantages in promoting experiential learning, though they also identified limitations related to content design and pedagogical alignment.

Work by Zhi and Wu~\citep{zhi2023extended} proposed a cognitive-affective model of immersive learning, arguing that XR-based language learning environments enhance both cognitive processing and emotional engagement, leading to deeper learning outcomes. Godwin-Jones~\citep{godwinJones2023presence} explored the notions of presence and agency in virtual spaces, highlighting the potential of XR for creating authentic language learning experiences where learners can exercise autonomy and make meaningful communicative choices. Panagiotidis~\citep{panagiotidis2021virtual} examined VR applications specifically designed for language learning, finding that virtual environments can effectively complement traditional instruction by offering novel modalities for practice and assessment.

Despite these advances, Taborda et al.~\citep{taborda2025engagement} noted that engagement and attention in XR learning environments remain under-investigated, with limited understanding of how immersive features affect sustained learning. Zhang et al.~\citep{zhang2023concepts} identified a need for stronger theoretical frameworks to steer the integration of mixed reality technologies into language curricula, arguing that without such frameworks, XR tool deployment risks being technology-driven rather than pedagogically grounded. Garcia et al.~\citep{garcia2024binding} explored the intersection of AI and XR in design education, identifying both challenges and opportunities that emerge when emerging technologies are applied in educational contexts.

\subsection{AI-Driven Speech Recognition and Translation in XR}
\label{sec:related_asr}

Advances in automatic speech recognition (ASR) have been accelerated by deep learning architectures, particularly encoder-decoder frameworks and transformer models~\citep{radford2023robust, nllb2022no}. OpenAI's Whisper model constitutes a significant milestone in ASR, attaining robust multilingual speech recognition through training on over 680{,}000 hours of weakly supervised audio data~\citep{radford2023robust}. Whisper's capacity to generalise across languages and acoustic conditions renders it well-suited for incorporation into XR environments where real-time, precise transcription is essential for inclusive communication~\citep{radford2023robust, hartholt2019ubiquitous}.

Multilingual translation has also witnessed substantial progress through models such as Meta's No Language Left Behind (NLLB), which supports translation across 200 languages using a conditional compute architecture based on Sparsely Gated Mixture of Experts~\citep{nllb2022no}. The NLLB model achieves a 44\% improvement in BLEU scores relative to prior state-of-the-art systems, with notable gains for low-resource languages~\citep{nllb2022no}. Within XR contexts, real-time translation services facilitate collaboration among multilingual users, effectively breaking down language barriers and enriching user interactions~\citep{sylaiou2023use, hirzle2023when}. Research has concentrated on bridging the modality gap between speech and text to ensure effective cross-modal communication within immersive settings~\citep{liu2020bridging}, while multilingual audio-visual corpora such as MuAViC have been constructed to support robust speech-to-text translation across modalities~\citep{anwar2023muavic}.

Hartholt et al.~\citep{hartholt2019ubiquitous} described a multi-platform framework for embodied AI agents in XR, demonstrating the potential of virtual humans to function as ubiquitous interaction partners delivering contextualised language instruction. Sylaiou et al.~\citep{sylaiou2023use} explored the use of XR technologies for enhancing visitor experience and inclusion at industrial museums, demonstrating how real-time transcription and translation can improve accessibility in cultural settings. Hirzle et al.~\citep{hirzle2023when} provided a scoping review of the intersection between XR and AI, charting the research landscape at this convergence and identifying key opportunities and challenges for future development.

\subsection{Sign Language Translation and Recognition}
\label{sec:related_sl}

AI has markedly advanced sign language translation and recognition, facilitating communication between deaf and hearing individuals by converting spoken or written language into sign language gestures~\citep{rodriguezcorrea2023benefits, camgoz2020sign}. Deep learning models, particularly transformer-based architectures, have yielded substantial improvements in both translation accuracy and efficiency~\citep{chaudhary2022signnet, zhou2023glossfree}. Current systems frequently employ gloss annotation, which decomposes signs into linguistic components to improve translation quality~\citep{camgoz2020sign}, while more recent gloss-free approaches have achieved comparable outcomes by eliminating this intermediate step~\citep{zhou2023glossfree}. Visual-language pretraining techniques have proven effective in enhancing both the accuracy and scalability of sign language translation, as demonstrated on benchmark datasets such as PHOENIX14T and CSL-Daily~\citep{camgoz2020sign, zheng2023cvtslr}. Contrastive visual-textual transformation approaches, such as CVT-SLR, have been proposed for sign language recognition with variational alignment, yielding strong results on standard benchmarks~\citep{zheng2023cvtslr}. Additionally, modern wearable and sensor-based technologies possess the potential to complement AI systems by enabling real-time gesture recognition and enhancing accessibility for individuals with hearing impairments~\citep{wu2023ultra}. Taborri et al.~\citep{taborri2023use} reviewed the use of AI for sign language recognition in education, with a focus on the ISENSE project, while Strobel et al.~\citep{strobel2023artificial} applied design science research to develop AI-based sign language translation systems.

Google MediaPipe has emerged as a powerful open-source framework for real-time hand and body tracking, providing 21 three-dimensional hand landmarks from individual video frames~\citep{lugaresi2019mediapipe}. Its lightweight architecture and ability to operate on mobile devices make it particularly suitable for integration with XR applications where computational efficiency is critical. MediaPipe has been widely adopted in sign language recognition research, enabling the extraction of precise hand and finger coordinates that serve as input features for gesture classification models~\citep{lugaresi2019mediapipe, subramanian2022integrated}. Despite these advances, significant challenges persist, particularly in developing comprehensive systems for International Sign (IS). Unlike national sign languages such as ASL or British Sign Language, IS is a visual communication system used by deaf individuals from diverse linguistic backgrounds in international contexts, such as conferences, sporting events, and educational settings~\citep{eud2018position, asltransformer}. Online sign language repositories such as SpreadTheSign~\citep{spreadthesign} have contributed to cross-linguistic accessibility by providing video demonstrations of signs across multiple national sign languages, serving as valuable training resources for computational models. Recent work by Srivastava et al.~\citep{srivastava2024continuous} demonstrated the effectiveness of MediaPipe Holistic combined with deep learning architectures for continuous sign language recognition, achieving promising results that underscore the potential of landmark-based approaches for real-time gesture classification. However, while some progress has been achieved with ASL translation models~\citep{handspeak}, advanced implementations for IS remain scarce, and the animation of sign language gestures via avatars in XR environments constitutes an open research challenge~\citep{yin2023gloss, tantaroudas2026xrsalento}.

\subsection{Large Language Models for Session Summarisation}
\label{sec:related_llm}

Large Language Models (LLMs) have expanded the capabilities of automatic summarisation by generating coherent and contextually appropriate summaries of complex data~\citep{ahmed2022fewshot, boros2023towards}. Models such as GPT-4 and fine-tuned variants of T5 and BART excel at processing large datasets and extracting pertinent information, making them well-suited for session summarisation in educational settings~\citep{boros2023towards}. Within XR environments, LLMs can synthesise immersive experiences by summarising key interactions, enabling learners to revisit essential insights from their sessions. The capacity of LLMs to perform abstractive summarisation allows them to generate coherent narratives rather than mere repetitions of content, thereby enhancing learning retention~\citep{boros2023towards}. Bozkir et al.~\citep{bozkir2024embedding} explored the embedding of LLMs into XR, identifying opportunities and challenges for inclusion, engagement, and privacy. Boros and Oyamada~\citep{boros2023towards} investigated LLM organisation for abstractive summarisation, while Ramprasad et al.~\citep{ramprasad2024analyzing} analysed LLM behaviour in dialogue summarisation, revealing trends in circumstantial hallucinations that can affect factual accuracy. Mitigating hallucination through fine-tuning and task-specific training remains a key area for solidifying the role of LLMs in adaptive learning systems within XR~\citep{ramprasad2024analyzing}.

\subsection{Sentiment Analysis in Educational Technology}
\label{sec:related_sentiment}

Sentiment analysis has gained traction as a means of enriching communication in educational technology by detecting and conveying emotional cues. Transformer-based models such as RoBERTa have exhibited strong performance in classifying textual inputs into emotional categories~\citep{barbieri2020tweeteval}. In the context of XR-based language learning, sentiment analysis can enhance avatar-mediated communication by mapping detected emotions to visual indicators, such as emoticons, thereby providing supplementary contextual information that improves emotional clarity and social presence for learners~\citep{barbieri2020tweeteval, tantaroudas2026xrsalento}.

% ============================================================
% 3. METHODOLOGY
% ============================================================
\section{Methodology}
\label{sec:methodology}

\subsection{Conceptual Framework and System Architecture}
\label{sec:arch}

\begin{figure}[H]
    \centering
    \includegraphics[width=0.65\textwidth]{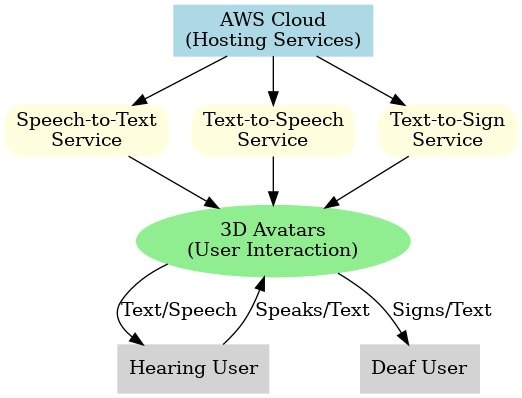}
    \caption{High-level overview of the proposed system architecture. AI services are hosted on AWS Cloud infrastructure and accessed by 3D avatars that deliver real-time language learning content through text, speech, and sign language modalities for both hearing and deaf users.}
    \label{fig:architecture}
\end{figure}

The proposed platform unifies modular AI-driven services designed for both deaf and hearing individuals within immersive XR environments. The system architecture, depicted in Figure~\ref{fig:architecture}, adopts a service-oriented approach in which each AI capability, speech-to-text, text-to-speech, text-to-text translation, text-to-sign translation, sentiment analysis, and session summarisation, is deployed as an independent microservice accessible through RESTful APIs. This modular design ensures flexibility, scalability, and interoperability across XR platforms. Additional details about the platform's communication capabilities are described in our companion publication~\citep{tantaroudas2026interact}.

The services are hosted on AWS Cloud infrastructure, ensuring high availability and the capacity to scale horizontally to handle growing user loads. In VR settings, 3D avatars serve as virtual tutors, delivering real-time language instruction through text and speech translation for hearing users and text-to-sign translation for deaf users. The XR scenarios were developed in consultation with educators and language acquisition specialists to ensure pedagogical relevance, aligning AI services with established communicative and task-based learning approaches.

To guarantee scalability, the platform employs Docker-based orchestration and cloud-native compatibility, enabling horizontal scaling of individual AI microservices. A load testing campaign simulated 1{,}000 concurrent users sending simultaneous API requests to the backend services, with the system maintaining an average response time under 800\,ms and registering no critical failures, confirming robust performance under high-demand educational scenarios.

\subsection{Speech-to-Text Transcription}
\label{sec:stt}

\begin{figure}[H]
    \centering
    \includegraphics[width=0.85\textwidth]{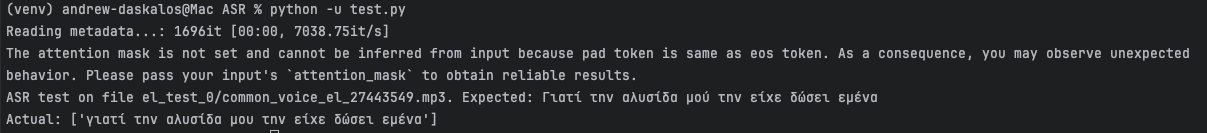}
    \caption{Speech-to-text transcription utilising a Whisper AI wrapper~\citep{radford2023robust}. The terminal output shows real-time transcription of Greek speech input with the expected and actual text outputs.}
    \label{fig:stt}
\end{figure}

The speech-to-text component leverages OpenAI's Whisper model~\citep{radford2023robust}, an open-source ASR system trained on over 680{,}000 hours of multilingual and multitask supervised data. Whisper employs a sequence-to-sequence transformer architecture that processes audio spectrograms and generates transcriptions, supporting robust performance across diverse languages and acoustic conditions. The platform deploys Whisper as an API service that receives audio input from users within the VR environment and returns real-time text transcriptions. Figure~\ref{fig:stt} illustrates the speech-to-text transcription process in Greek, demonstrating accurate conversion of spoken inputs into text. The terminal output displays real-time transcription of Greek speech input, demonstrating precise processing of spoken language into text for subsequent translation and presentation within the XR environment.

\subsection{Multilingual Text-to-Text Translation}
\label{sec:translation}

\begin{figure}[H]
    \centering
    \includegraphics[width=0.85\textwidth]{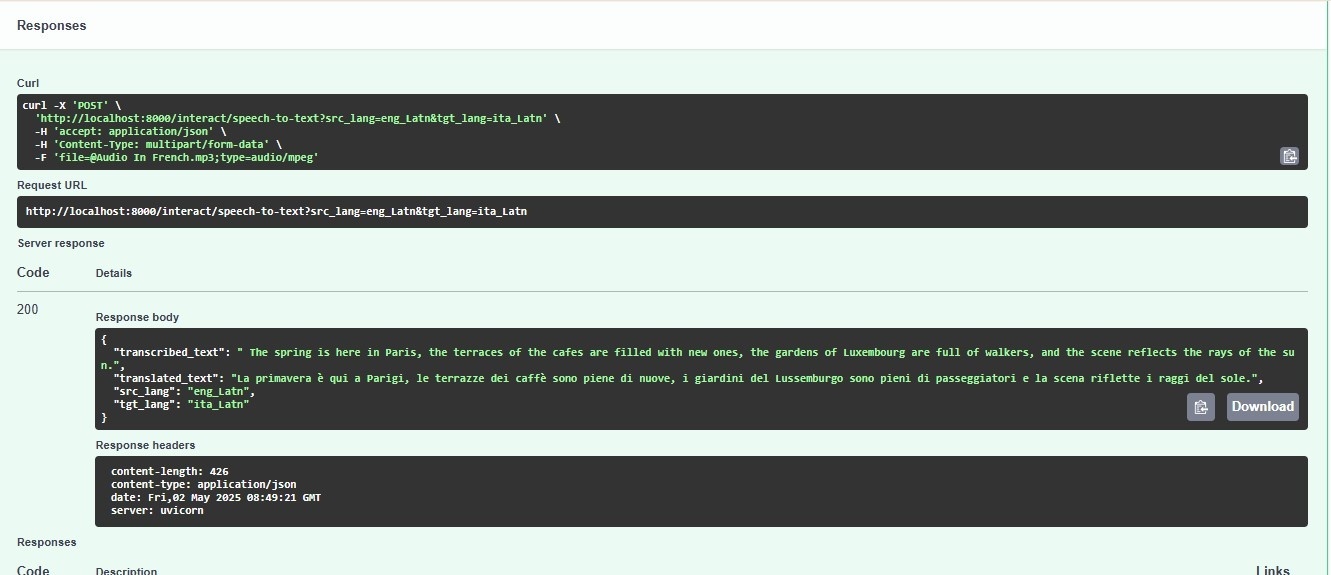}
    \caption{Text-to-text translation workflow employing Meta's NLLB model~\citep{nllb2022no}. The API interface shows a curl request with source and target language parameters, along with the server response containing both transcribed and translated text.}
    \label{fig:translation}
\end{figure}

For multilingual translation, the platform incorporates Meta's No Language Left Behind (NLLB) model~\citep{nllb2022no}, a conditional compute model based on Sparsely Gated Mixture of Experts that supports translation across 200 languages. The NLLB-200-distilled-600M variant is deployed as a translation API service, enabling real-time conversion of transcribed text between multiple language pairs. The translation service is chained sequentially with the speech-to-text service, establishing an automated pipeline in which user speech is first transcribed into text and subsequently translated into the target language chosen by the user. Figure~\ref{fig:translation} demonstrates the text-to-text translation workflow using NLLB, showcasing real-time translation across multiple languages. The system translates text in real time across multiple languages, enabling multilingual interactions within the XR learning environment.

\subsection{Text-to-Speech Synthesis}
\label{sec:tts}

The text-to-speech (TTS) component produces natural, real-time audio output in the target language. Initially, the platform explored the Coqui TTS Tacotron2-DDC model~\citep{coquitts} and the open-source Piper TTS system~\citep{pipertts}. However, owing to security vulnerabilities associated with poorly maintained Python dependencies in Piper and limitations in voice naturalness with Tacotron2-DDC, the team transitioned to AWS Polly, a production-grade TTS service that provides low-latency, high-quality speech synthesis across 34 languages.

The selection of AWS Polly was informed by a comprehensive benchmarking study comparing four TTS services: AWS Polly Standard, Google Cloud TTS Standard, Microsoft Azure Speech, and ElevenLabs. Table~\ref{tab:tts_latency} presents the latency performance metrics, derived from both the Picovoice TTS Latency Benchmark study~\citep{picovoice2024tts} and our own testing. AWS Polly was selected for its consistently low first-byte latency (50--100\,ms), cost-effectiveness (\$4 per million characters), and stable performance across testing sessions, avoiding the high variability observed with other services. The TTS service is integrated with the translation pipeline, providing multilingual spoken feedback through the 3D avatar system.

\begin{table}[H]
\centering
\caption{Latency performance metrics for text-to-speech services.}
\label{tab:tts_latency}
\begin{tabular}{lccc}
\toprule
\textbf{Service} & \textbf{First Byte Latency} & \textbf{FTTS} & \textbf{Total Response Time} \\
\midrule
AWS Polly Standard    & 50--100\,ms   & 450\,ms  & 780\,ms  \\
Google Cloud Standard & 300--2000\,ms & 600\,ms  & 1200\,ms \\
Microsoft Azure       & 150--800\,ms  & 1140\,ms & 1500\,ms \\
ElevenLabs            & 300--500\,ms  & 840\,ms  & 1250\,ms \\
\bottomrule
\end{tabular}
\end{table}

\subsection{Sentiment Analysis with Emoticon Mapping}
\label{sec:sentiment}

\begin{figure}[H]
    \centering
    \includegraphics[width=0.75\textwidth]{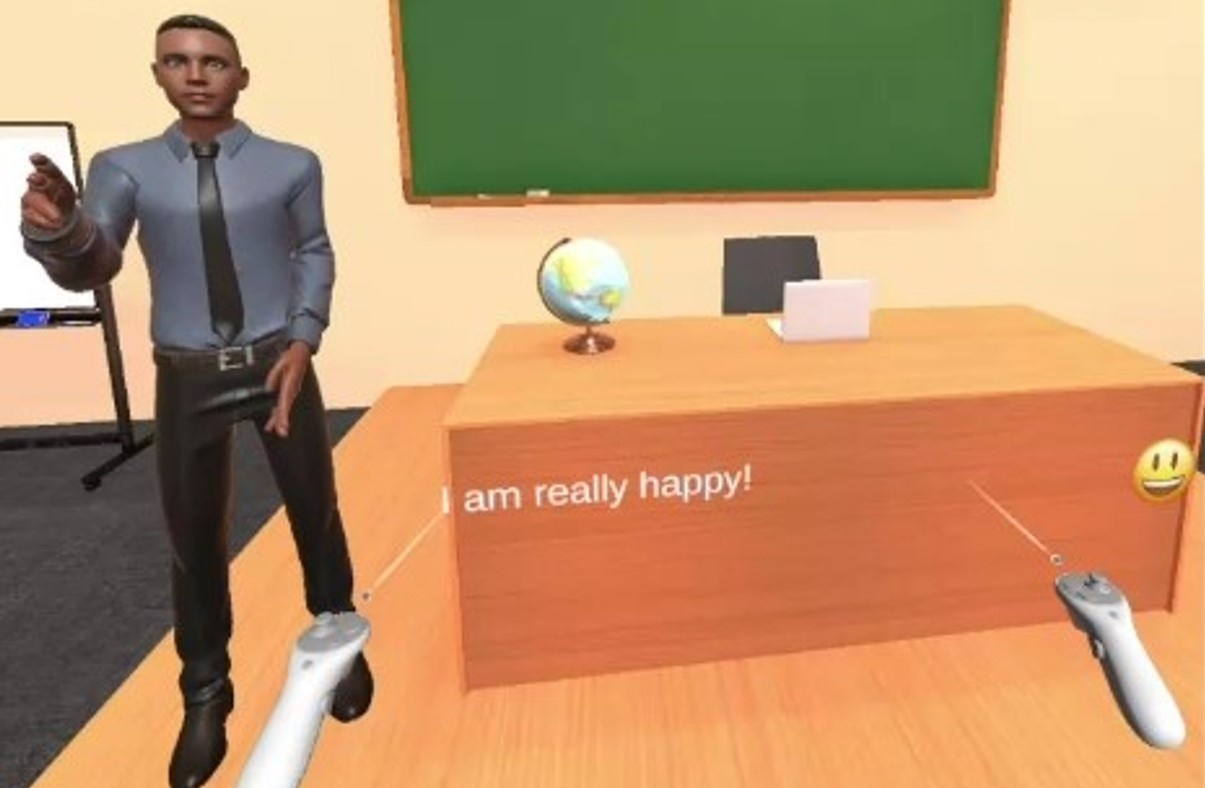}
    \caption{Emoticon-based sentiment feedback in the XR environment. The avatar delivers speech output (``I am really happy!'') while a corresponding emoticon is displayed to convey the detected emotional tone.}
    \label{fig:sentiment_vr}
\end{figure}

\begin{figure}[H]
    \centering
    \includegraphics[width=0.75\textwidth]{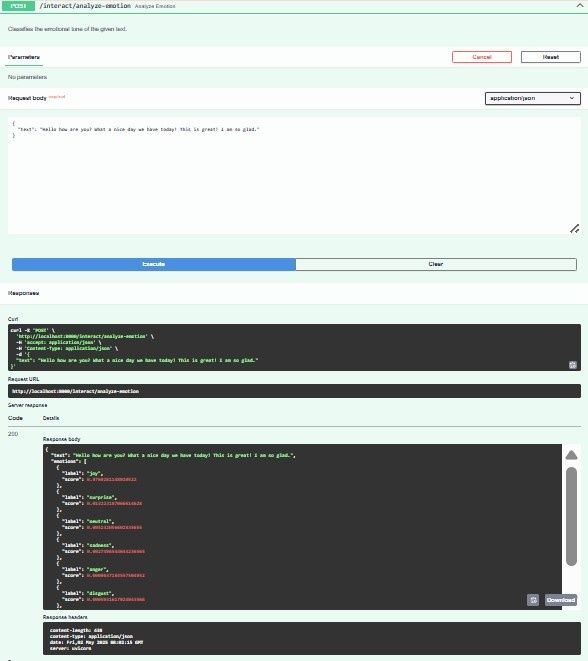}
    \caption{API request and response format for the RoBERTa-based sentiment analysis module. The interface accepts text input and returns classified emotion labels (joy, surprise, neutral, sadness, anger, disgust) with associated confidence scores.}
    \label{fig:sentiment_api}
\end{figure}

To enhance avatar-mediated communication with emotional context, the platform integrates a sentiment analysis module based on the twitter-roberta-base-sentiment model from CardiffNLP~\citep{barbieri2020tweeteval}. This RoBERTa-based transformer processes translated text input and classifies it into sentiment categories such as ``happy,'' ``neutral,'' or ``sad.'' Each detected sentiment is mapped to a corresponding emoticon that is displayed in real time within the VR scene adjacent to the avatar, augmenting affective expressiveness and social presence. Figure~\ref{fig:sentiment_vr} demonstrates the emoticon-based sentiment feedback displayed in the XR environment. The RoBERTa-based sentiment classifier processes the avatar's speech output and maps detected emotions to visual emoticons displayed alongside the avatar, providing supplementary contextual and emotional cues for learners. Figure~\ref{fig:sentiment_api} shows the API request and response format for the sentiment analysis module, illustrating the RESTful interface through which the VR application communicates with the sentiment service. The interface accepts JSON-formatted text input and returns classified sentiment labels with associated confidence scores for multiple emotional categories.

\subsection{Meeting Summarisation Module}
\label{sec:summarisation}

\begin{figure}[H]
    \centering
    \includegraphics[width=0.85\textwidth]{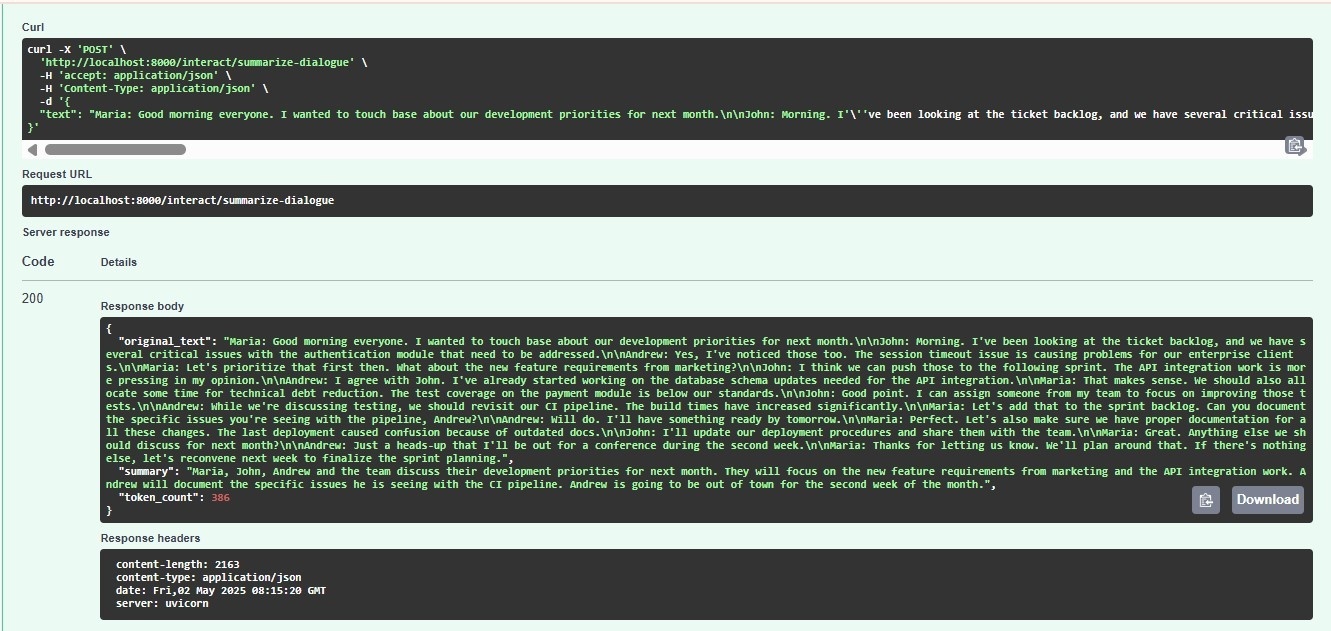}
    \caption{API request and response format for the meeting summarisation module based on the flan-t5-base-samsum model~\citep{schmid2022flant5}. The response includes the original dialogue text, a generated summary, and token count statistics.}
    \label{fig:summarisation}
\end{figure}

The platform integrates the flan-t5-base-samsum model~\citep{schmid2022flant5}, available on Hugging Face, for real-time summarisation of dialogues in multilingual XR-based educational scenarios. This model is deployed as an API on AWS Lambda as part of the platform's backend services. It is designed for deaf or hard-of-hearing users, interpreters, and educators to receive concise, natural language summaries of verbal exchanges within the XR scene. Figure~\ref{fig:summarisation} presents the API interface for the summarisation module, showing how input dialogue is processed and condensed into a coherent summary. The system processes dialogue text and generates succinct summaries, with the response including both the original text and the summarised output along with token count statistics.

\subsection{International Sign (IS) Translation}
\label{sec:is_translation}

A central contribution of this work is the development of an IS translation pipeline. During the research phase, a thorough review revealed that while each country possesses its own distinct sign language with unique linguistic structures, IS functions as a broadly understood visual communication system used by deaf individuals from diverse linguistic backgrounds in international contexts~\citep{eud2018position, asltransformer}. Unlike national sign languages, IS draws upon signs from multiple sign languages, iconic gestures, and universal visual cues to enable understanding across national boundaries.

To develop the IS translation model, approximately 750 videos of IS gestures were collected and processed using Google MediaPipe~\citep{lugaresi2019mediapipe} and OpenCV~\citep{bradski2008learning}, extracting key movement coordinates and hand position data from 21 three-dimensional hand landmarks per frame. The MediaPipe Hands solution provides a machine learning pipeline that infers hand landmarks from individual frames, outputting 21 key points per hand with $x$, $y$, and $z$ coordinates normalised to the image frame. For each gesture video, frames were extracted at a uniform sampling rate and processed through the MediaPipe pipeline to obtain temporal sequences of landmark positions. These sequences were subsequently normalised relative to the wrist landmark to account for variations in hand size, camera distance, and signer morphology. The normalised landmark sequences were aggregated into a structured dataset associating each sequence with its corresponding IS sign label. Resources such as HandSpeak~\citep{handspeak} and SpreadTheSign~\citep{spreadthesign} provided reference video demonstrations of IS signs, which were used both for dataset curation and validation of sign-to-label mappings.

\begin{figure}[H]
    \centering
    \includegraphics[width=0.75\textwidth]{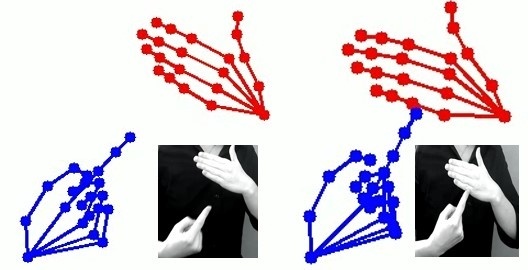}
    \caption{Visual sequence of real-time avatar animation driven by extracted gesture landmarks from the IS dataset. Left panels: original hand gesture videos with overlaid MediaPipe landmarks. Right panels: corresponding 3D avatar performing the recognised signs.}
    \label{fig:is_gestures}
\end{figure}

This dataset served as the basis for training a gesture classification model that maps text inputs to corresponding IS signs. An API was developed to map the classified hand positions and gestures to 3D avatar animations within the Unity-based VR environment, enabling real-time IS interpretation. The avatar animation system translates the classified landmark sequences into joint rotations applied to the avatar's skeletal rig, with interpolation between keyframes to ensure smooth transitions. Preliminary evaluations indicate that avatar animation latency consistently remains under 300\,ms, ensuring natural, real-time communication for XR users. Figure~\ref{fig:is_gestures} presents a visual sequence of the real-time avatar animation pipeline, showing how extracted gesture landmarks are translated into avatar movements. The left panels display the original hand gesture videos with overlaid MediaPipe landmarks (in red and blue), while the right panels illustrate the corresponding 3D avatar performing the recognised sign within the VR environment.

\subsection{Immersive VR Learning Environment}
\label{sec:vr_env}

The final application was constructed using the Unity game engine and deployed on Meta Quest~3 headsets, providing a fully immersive learning experience. The Unity development environment was chosen for its cross-platform compatibility, extensive asset ecosystem, and native support for XR development through the XR Interaction Toolkit and OpenXR standards. The Meta Quest~3 headset was selected as the target deployment platform owing to its standalone operation (requiring no tethered PC), high-resolution passthrough capabilities for potential AR extensions, and growing adoption in educational and enterprise settings.

\begin{figure}[H]
    \centering
    \includegraphics[width=0.75\textwidth]{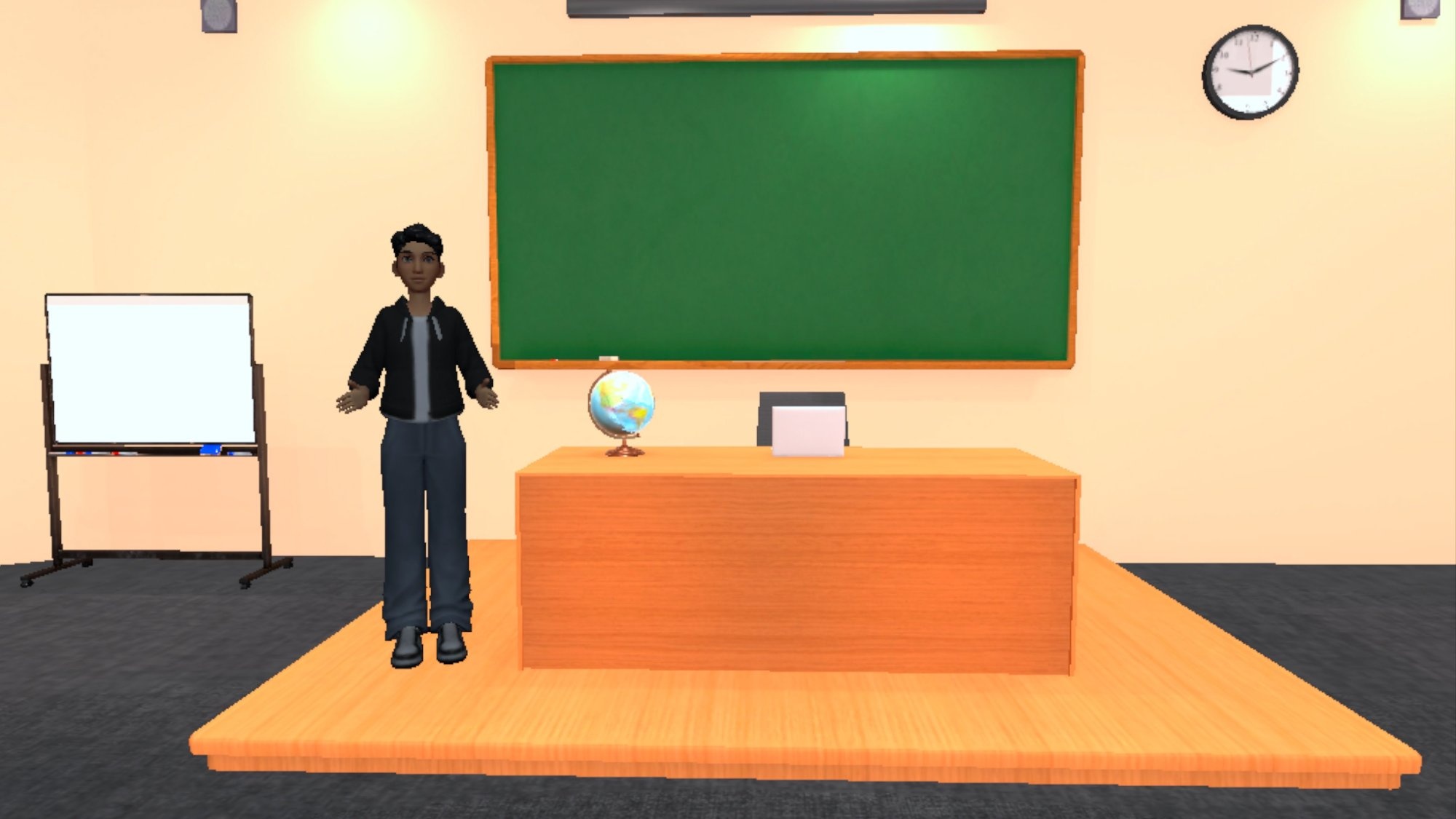}
    \caption{Immersive VR-based language learning system. A 3D avatar stands in a virtual classroom, delivering educational content through speech, text, and International Sign translation within a multilingual VR setting.}
    \label{fig:vr_classroom}
\end{figure}

Figure~\ref{fig:vr_classroom} illustrates the immersive VR classroom environment where a 3D avatar stands in a virtual classroom equipped with multilingual AI tools. Users interact with the system by selecting their preferred language through an interactive questionnaire presented within the VR interface; in response, the avatar provides real-time translations in both spoken language and IS-based sign translations of selected words and phrases. The avatar's animation system supports lip-sync with generated speech output and gestural animation for IS signs, with blending between idle, speaking, and signing states managed through Unity's Animator Controller.

This environment demonstrates how speech-to-text, text-to-sign translation, and text-to-speech services converge to create equitable, self-directed learning opportunities for both deaf and hearing users. The AI-powered avatar delivers educational content in a simulated classroom environment, translating spoken language into International Sign (IS) within a multilingual VR setting. The environment is designed for equitable access and cross-linguistic interaction, with all AI services hosted on scalable AWS Cloud infrastructure.

\begin{figure}[H]
    \centering
    \includegraphics[width=0.75\textwidth]{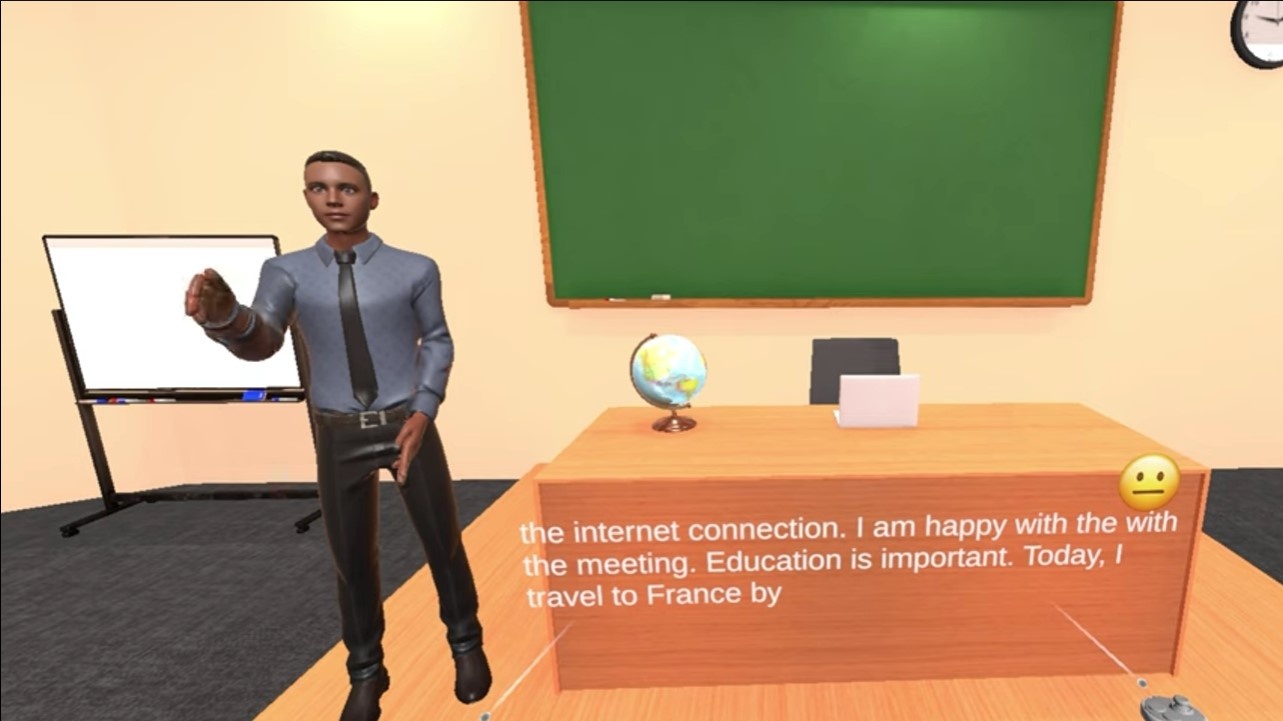}
    \caption{MVP demonstration showing multilingual avatar interaction in VR. The avatar delivers IS signs while text panels display transcription and translation output, alongside sentiment-based emoticon feedback.}
    \label{fig:mvp}
\end{figure}

Figure~\ref{fig:mvp} shows a snapshot of the pipeline from voice input to sign and spoken output in the Meta Quest~3 environment. The avatar is shown delivering IS signs in the virtual classroom environment, with text display panels visible for transcription and translation output.

% ============================================================
% 4. RESULTS
% ============================================================
\section{Results}
\label{sec:results}

\subsection{System Implementation and Integration}
\label{sec:results_system}

The proposed platform successfully integrates six AI-driven services into a cohesive XR learning experience. Table~\ref{tab:services} summarises the implemented services, their underlying models, and key performance characteristics. A comprehensive description of the full platform in the context of inclusive communication is available in~\citep{tantaroudas2026interact, tantaroudas2026aiservices}.

\begin{table}[H]
\centering
\caption{Summary of AI services implemented in the proposed platform.}
\label{tab:services}
\small
\begin{tabularx}{\textwidth}{lXX}
\toprule
\textbf{Service} & \textbf{Model/Technology} & \textbf{Key Characteristic} \\
\midrule
Speech-to-Text       & OpenAI Whisper~\citep{radford2023robust}          & Multilingual ASR, 680k+ hours training data \\
Text-to-Text Transl. & Meta NLLB-200~\citep{nllb2022no}                  & 200 languages, Mixture of Experts architecture \\
Text-to-Speech       & AWS Polly                                          & 34 languages, low-latency synthesis \\
Sentiment Analysis   & RoBERTa~\citep{barbieri2020tweeteval}              & Multi-class emotion classification \\
Session Summar.      & flan-t5-base-samsum~\citep{schmid2022flant5}       & Abstractive dialogue summarisation \\
IS Translation       & MediaPipe + Avatar~\citep{lugaresi2019mediapipe}   & 750 gesture videos, $<$300\,ms latency \\
\bottomrule
\end{tabularx}
\end{table}

\subsection{TTS Benchmarking}
\label{sec:results_tts}

Beyond the latency analysis presented in Table~\ref{tab:tts_latency}, voice quality metrics were also evaluated. Table~\ref{tab:tts_quality} presents the Mean Opinion Score (MOS) and Word Error Rate (WER) values reported by service providers for the compared TTS services. Table~\ref{tab:tts_capabilities} summarises the service capabilities, which represent important considerations for ensuring the platform remains accessible. Based on the benchmarking, AWS Polly Standard was selected for the proposed platform owing to its lowest and most consistent first-byte latency (50--100\,ms), cost-effectiveness (\$4 per million characters), and stable performance across testing sessions, avoiding the high variability observed with alternative services.

\begin{table}[H]
\centering
\caption{Voice quality metrics for text-to-speech services.}
\label{tab:tts_quality}
\begin{tabular}{lcc}
\toprule
\textbf{Service} & \textbf{Mean Opinion Score (MOS)} & \textbf{Word Error Rate} \\
\midrule
AWS Polly Standard    & 3.5--3.8  & 4.2      \\
Google Cloud Standard & 3.2--3.5  & Variable \\
Microsoft Azure       & 3.8--4.0  & 3.0      \\
ElevenLabs            & 3.83--4.2 & 2.83     \\
\bottomrule
\end{tabular}
\end{table}

\begin{table}[H]
\centering
\caption{Service capabilities of TTS services (standard models).}
\label{tab:tts_capabilities}
\begin{tabular}{lccc}
\toprule
\textbf{Service} & \textbf{Languages} & \textbf{Voices} & \textbf{Max Length} \\
\midrule
AWS Polly Standard    & 34   & 66    & 3000 \\
Google Cloud Standard & 40+  & 220+  & 5000 \\
Microsoft Azure       & 140+ & 110+  & 5000 \\
ElevenLabs            & 29   & 5000+ & 5000 \\
\bottomrule
\end{tabular}
\end{table}

\subsection{Benchmarking NLLB against EuroLLM}
\label{sec:results_benchmark}

To evaluate potential enhancements to the translation component, a comprehensive benchmark comparison was conducted between the deployed NLLB-200-distilled-600M model and two variants of the EuroLLM 1.7B model. All experiments were performed on a consumer-grade workstation equipped with an NVIDIA GeForce RTX 4060 GPU (8\,GB VRAM), ensuring that the benchmarking conditions reflect realistic deployment scenarios rather than high-end server infrastructure. The benchmarking utilised a test dataset of 10 English-to-French translations with varying complexity levels: simple conversational phrases (3 examples), medium-complexity technical sentences (4 examples), and complex sentences with specialised terminology (3 examples). Models were loaded sequentially to manage the limited GPU memory, and inference was conducted using float16 precision to maximise throughput within the available VRAM. For each model, translation quality (BLEU scores), inference speed, and resource utilisation were measured. Table~\ref{tab:benchmark} presents the comparative results.

\begin{table}[H]
\centering
\caption{NLLB vs.\ EuroLLM performance comparison.}
\label{tab:benchmark}
\small
\begin{tabularx}{\textwidth}{Xccc}
\toprule
\textbf{Metric} & \textbf{NLLB-200} & \textbf{EuroLLM 1.7B Base} & \textbf{EuroLLM 1.7B Inst.} \\
\midrule
Average BLEU Score       & 79.25     & 27.58     & 84.34     \\
Avg.\ Transl.\ Time (s) & 0.596     & 1.509     & 0.529     \\
Model Load Time (s)      & 26.63     & 25.37     & 40.99     \\
Memory Usage (GB)        & $\sim$2.5 & $\sim$3.5 & $\sim$3.5 \\
Successful Translations  & 10/10     & 10/10     & 10/10     \\
\bottomrule
\end{tabularx}
\end{table}

The principal findings from the benchmarking are as follows. The EuroLLM 1.7B Instruct model attained the highest BLEU score (84.34), surpassing NLLB (79.25) by approximately 5 points, demonstrating superior translation quality for European language pairs. However, the EuroLLM Base model performed poorly (27.58), underscoring the critical importance of instruction tuning for translation tasks. Regarding inference speed, EuroLLM 1.7B Instruct demonstrated the fastest average translation time (0.529\,s), marginally faster than NLLB (0.596\,s), while the base model was considerably slower at 1.509\,s per translation. Purpose-built sequence-to-sequence models (NLLB) exhibited strong performance, while causal LM models averaged 61.37 BLEU across all variants. The EuroLLM models require more memory ($\sim$3.5\,GB vs.\ $\sim$2.5\,GB for NLLB's distilled version), with longer initial loading times for the instruction-tuned variant. The results present a compelling case for considering EuroLLM 1.7B Instruct as an alternative to NLLB in the translation pipeline, particularly given its specialised focus on European languages that aligns with the proposed platform's target markets.

% ============================================================
% 5. DISCUSSION
% ============================================================
\section{Discussion}
\label{sec:discussion}

The findings demonstrate that the proposed platform successfully brings together multiple AI-driven services into a unified XR learning environment that addresses both the communication needs of hearing users and the accessibility requirements of deaf users. The modular, service-oriented architecture enables independent scaling and updating of individual components, which is essential for maintaining system performance as user demands increase and AI models evolve.

The successful unification of six AI services, speech-to-text, text-to-text translation, text-to-speech, sentiment analysis, session summarisation, and IS translation, within a single XR environment validates the premise that combining multiple AI modalities within immersive settings can generate meaningful, accessible learning experiences. The IS-focused approach, leveraging International Sign Language as a lingua franca rather than a single national sign language, holds the potential to broaden accessibility beyond national sign language boundaries and serve deaf learners from diverse linguistic backgrounds. Further discussion of the platform's applicability to inclusive communication scenarios, including business meeting contexts, is provided in~\citep{tantaroudas2026interact, tantaroudas2026business}.

The avatar-mediated sign language delivery, while achieving animation latency under 300\,ms, represents an initial proof of concept that would benefit from additional refinement. In particular, the quality of non-manual markers (facial expressions, head movements, body posture) is acknowledged in the broader literature as essential for comprehensible and natural signing~\citep{chaudhary2022signnet, yin2023gloss}, and their incorporation into the avatar system represents a significant area for future development.

The benchmarking analyses furnish quantitative evidence for technology selection decisions. The TTS evaluation confirmed that AWS Polly offers the best balance of latency, cost, and consistency for real-time VR applications, while the NLLB vs.\ EuroLLM comparison revealed that instruction-tuned causal language models can outperform purpose-built translation models in both quality and speed for European language pairs. This finding carries implications not only for the proposed platform but also for the broader UTTER project's development of European language models.

Several limitations warrant acknowledgement. The IS dataset of 750 gesture videos, while sufficient for an initial proof-of-concept, represents a limited vocabulary that constrains the expressiveness of the translation system. Additionally, formal usability studies with standardised instruments (e.g., System Usability Scale, NASA-TLX for cognitive load) have not yet been conducted and represent an important direction for future validation. The system has not yet been evaluated with end users wearing VR headsets in naturalistic learning scenarios, which is necessary to assess immersion, spatial presence, and actual learning outcomes.

The ethical dimensions of the platform were addressed through adherence to GDPR regulations, anonymisation of voice and gesture data, secure API access controls, and the application of FAIR data principles. Efforts were made to ensure fairness across languages and cultural contexts, particularly in the design and training of sign language avatars, to avoid misrepresentation and bias.

The platform presented in this study addresses a key gap identified in the literature: the absence of integrated, multimodal AI systems within XR that serve both deaf and hearing learners simultaneously. While previous work has typically focused on individual AI capabilities in isolation, such as speech recognition for captioning~\citep{radford2023robust} or sign language recognition for classification~\citep{chaudhary2022signnet, srivastava2024continuous}, our system demonstrates the feasibility of orchestrating multiple AI services into a unified educational experience. This approach aligns with the vision articulated by Hirzle et al.~\citep{hirzle2023when}, who identified the convergence of XR and AI as a high-potential research frontier, and extends it by grounding the integration in a concrete, deployable platform validated through comprehensive technical benchmarking. A detailed account of the platform's design and its application to inclusive language learning is provided in~\citep{tantaroudas2026aiservices}.

From a policy perspective, the platform's focus on IS as a lingua franca for deaf communication across national boundaries aligns with the objectives of the European Accessibility Act and the EU's commitment to digital inclusion. By providing an XR-based learning environment that supports both spoken multilingual interaction and sign language translation, the proposed platform contributes to the broader agenda of equitable access to education, as articulated in the EU Digital Education Action Plan (2021--2027). The modular, cloud-native architecture ensures that the platform can be adapted to diverse institutional contexts, from formal language schools to informal community-based learning settings. Furthermore, the same underlying AI services have demonstrated applicability beyond language education, including accessible communication in professional and business meeting contexts~\citep{tantaroudas2026business}.

% ============================================================
% 6. CONCLUSIONS
% ============================================================
\section{Conclusions}
\label{sec:conclusions}

This paper presented a comprehensive framework for integrating modular AI-driven services into immersive XR environments to support language education for both deaf and hearing individuals. The platform combines speech-to-text transcription, multilingual translation, text-to-speech synthesis, sentiment analysis, and International Sign translation, all delivered through AI-powered 3D avatars within a Unity-based VR environment deployed on Meta Quest~3 headsets.

The system's modular architecture and successful integration of all AI components demonstrate the technical feasibility of the approach. Quantitative benchmarking of TTS services and multilingual translation models provided evidence-based justification for technology selection and identified promising alternatives for future integration.

Future work will concentrate on several priorities. First, the IS vocabulary will be substantially expanded by collecting and processing additional gesture videos from diverse interpreters, with improved gesture landmark normalisation to ensure smoother and more consistent avatar animations across signers. Second, avatar facial expressions and lip-sync functionality will be developed to provide the non-manual markers that are essential for natural sign language communication. Third, the single-user experience will be extended into a multiplayer VR setting where hearing and deaf users can communicate in real time, better showcasing the integrated sentiment analysis and translation capabilities and enabling collaborative language learning scenarios. Fourth, formal user studies with larger and more diverse participant groups will be designed and conducted, incorporating standardised usability instruments (e.g., System Usability Scale, NASA-TLX), pre/post learning assessments, and cognitive load measurements to produce quantitative evidence of learning effectiveness. Fifth, the EuroLLM 1.7B Instruct model will be further evaluated for integration into the production translation pipeline, with expanded benchmarking across additional European language pairs and domain-specific terminology. Sixth, the platform architecture will be extended to support additional XR hardware beyond Meta Quest~3, including standalone AR devices and desktop-based VR systems, to maximise accessibility across institutional contexts. Finally, the platform will be piloted in formal educational settings, including schools and language training centres, to assess real-world adoption, learning outcomes, and alignment with the EU Digital Education Action Plan's goals for inclusive, technology-enhanced learning.

% ============================================================
% GRANT INFORMATION
% ============================================================
\section*{Grant Information}
This research was supported by FSTP Funding from the European Union's Horizon Europe Research and Innovation programme under grant agreement No.\ 101070631 (UTTER, Unified Transcription and Translation for Extended Reality) and from the UK Research and Innovation (UKRI) under the UK government's Horizon Europe funding guarantee (Grant No.\ 10039436). Views and opinions expressed are those of the authors only and do not necessarily reflect those of the European Union nor the granting authority. Neither the European Union nor the granting authority can be held responsible for them. The funding body had no role in study design, data collection and analysis, decision to publish, or preparation of the manuscript.

% ============================================================
% ACKNOWLEDGEMENTS
% ============================================================
\section*{Acknowledgements}
The authors gratefully acknowledge the UTTER consortium for the provision of the FSTP funding mechanism enabling the development of this research. The UTTER project partners, Maryam Hashemi (University of Amsterdam) and Amin Farajian (UNBABEL), are acknowledged for their guidance during the implementation of this study.

% ============================================================
% REFERENCES
% ============================================================
\bibliographystyle{unsrtnat}
\bibliography{references}

\end{document}